\documentclass[aip,rsi,floatfix,preprint]{revtex4-1}

\usepackage{graphicx}

\begin{document}

\title{\Huge A high sensitivity ultra-low temperature RF conductance and noise measurement setup}

\author{F. D. Parmentier$^{1, \ddag}$, A. Mah\'e$^{1}$, A. Denis$^{1}$, J.-M. Berroir$^{1}$, D.C. Glattli$^{1, \dag}$, B. Pla\c cais$^{1}$, G. F\`eve$^{1}$} 

\affiliation{$^{1}$ Laboratoire Pierre Aigrain, Ecole Normale Sup\'erieure, CNRS (UMR 8551), Universit\'{e} P. et M. Curie, Universit\'{e} D. Diderot\\
24, rue Lhomond, 75231 Paris Cedex 05, France.
}

\date{\today}
\maketitle

\textbf{We report on the realization of a high sensitivity RF noise measurement scheme to study small current fluctuations of mesoscopic systems at milliKelvin temperatures. The setup relies on the combination of an interferometric amplification scheme and a quarter-wave impedance transformer, allowing the measurement of noise power spectral densities with GHz bandwidth up to five orders of magnitude below the amplifier noise floor. We simultaneously measure the high frequency conductance of the sample by derivating a portion of the signal to a microwave homodyne detection. We describe the principle of the setup, as well as its implementation and calibration. Finally, we show that our setup allows to fully characterize a subnanosecond  on-demand single electron source. More generally, its sensitivity and bandwidth make it suitable for applications manipulating single charges at GHz frequencies.}

\section{Introduction}

Many informations on electronic properties of mesoscopic systems have been obtained through studies of electronic noise\cite{BlanterButtiker,DelattreKondo,Marcus,FermionicHBT,KumarQPC,HerrmannPRL,SaminadayarPRL,DePiciottoNature}. Measurements of small current fluctuations (typically a few $10^{-29}A^{2}/Hz$) at low frequency have put into light spectacular phenomena in mesoscopic conductors, such as the suppression of shot noise\cite{KumarQPC,HerrmannPRL} which demonstrates that the Pauli exclusion principle correlates the flow of electrons participating in mesoscopic currents, or the fractional charge in 2D electron systems\cite{SaminadayarPRL,DePiciottoNature}. Low frequency noise measurements have since become common experimental techniques \cite{GlattliJAP,MarcusRSI,GlattliHouches}. Comparatively, studies of current fluctuations at microwave frequencies \cite{ReznikovPRL,SchoelkopfPRL,GabelliPRL,Portier,Hakonengraphene,EZB,Gross2010,Gross2010B} are much less widespread. However, high frequency noise is highly relevant to probe basic phenomena such as electron/photon statistics in quantum conductors\cite{GabelliPRL,EZB}, and will prove useful to characterize systems manipulating single electrons at GHz frequencies\cite{Fevescience, Blumenthalnature}. In this context, the small magnitude of current fluctuations, typically $S_{i}\propto e^2f \approx 2.5\times10^{-29}A^{2}/Hz$ at $f=1GHz$, makes its measurement very challenging, especially as fast single charge detection suffers a mismatch problem between the high impedance ($Z\propto h/e^2\approx26k\Omega$) of quantum sources and the low ($50\Omega$) impedance of microwave amplifiers. This can hardly be overcome in broadband high-frequency experiments and strongly alters the current noise power resolution (by typically five orders of magnitude), which can only be recovered by increasing the measuring time. 
A standard RF noise measurement method consists in integrating the noise power spectral density over a finite bandwidth using square law detectors (see Fig.~\ref{SNR}a). One has to take into account the noise of the first amplifier in the setup, which is usually significantly larger than the noise of the sample. The resolution is limited by the integration time, which becomes very large and may eventually exceed the timescale over which the amplification gain can vary, thus making the measurement method less reliable.

In this paper, we present a highly sensitive, wideband microwave frequency noise measurement technique with a current noise resolution lying an order of magnitude below the $e^2f$ threshold. We have used the implemented setup to study the current fluctuations emitted by a single electron source\cite{Fevescience, MaheJLTP}, which consists of a mesoscopic capacitor \cite{Gabelliscience} driven at microwave frequencies. The coupling between the source and the amplifiers is first increased by using a broad-band $120\Omega$ to $50\Omega$  quarter-wave impedance transformer. The signal is then amplified with a phase-modulated double balanced amplifier. This setup allows a highly stable amplification on a broad bandwidth ($1.2-1.8GHz$) of very low signals emitted at the base temperature of a dilution refrigerator.

In the first part of the article, we recall the principle of the modulated double balanced amplifier, and its advantages compared to a direct amplification technique. We also describe its implementation, including a microwave homodyne detection of the conductance, inside an Oxford Kelvinox 400 dilution refrigerator, as well as its calibration using Johnson noise thermometry. In the second part, we describe the impedance transformer and its realization inside a sample holder connected to the mixing chamber of the dilution refrigerator. We finally present a typical operation of the whole setup, demonstrating a stability of the measurement over more than 40 hours and a sensitivity of about $2\times10^{-28}A^{2}/Hz/\sqrt{Hz}$ ($1.15\times10^{-29}A^{2}/Hz$ in a 5 minutes integration time).

\section{Modulated double balanced amplifier}
\begin{figure*}[!htph]
\centering\includegraphics[width=0.95\textwidth]{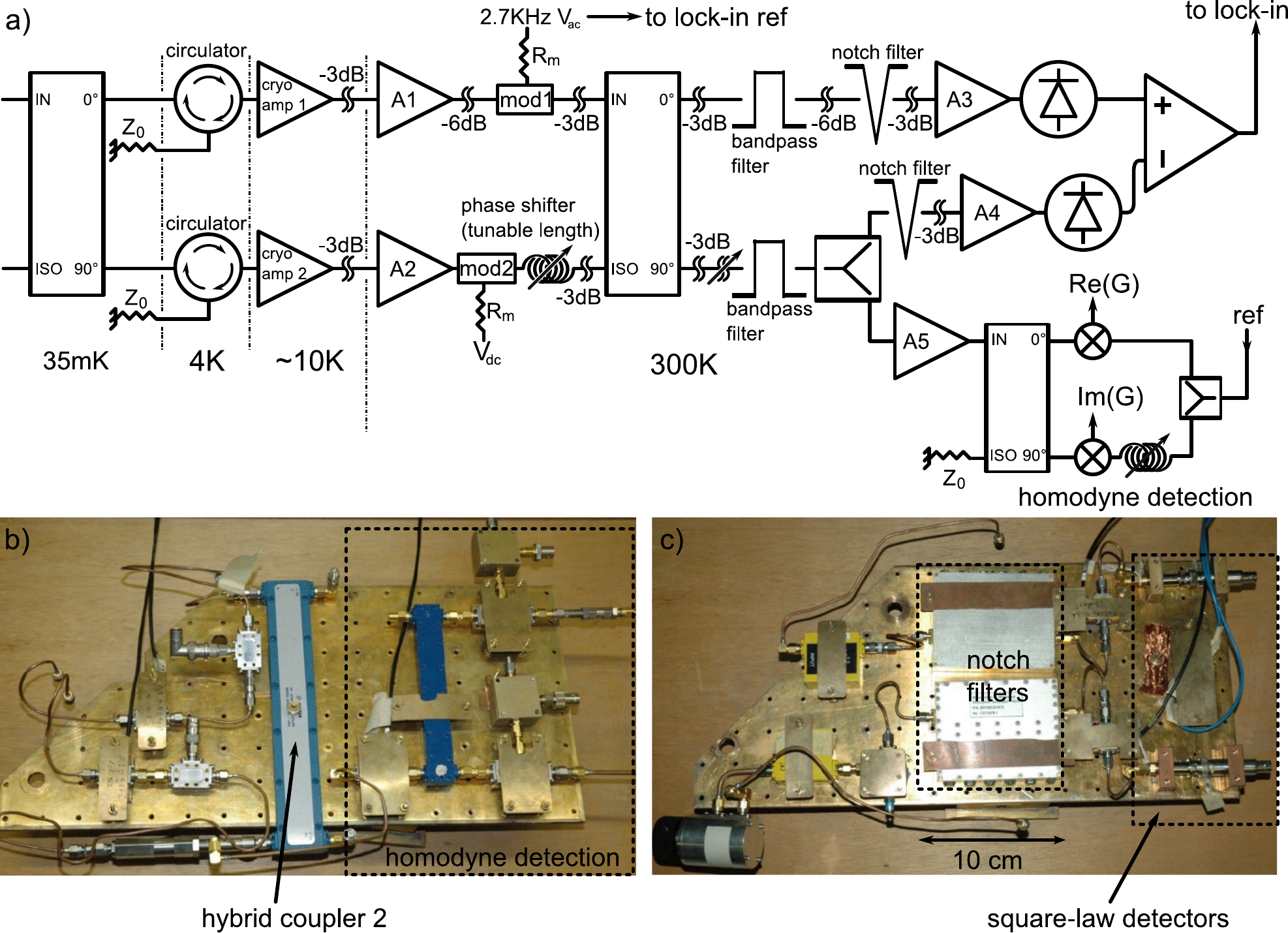}
\caption{a) Schematic of the setup, as implemented in our Oxford Kelvinox 400 dilution refrigerator. b) and c) Pictures of the room-temperature parts of the setup.
}
\label{expsetup}
\end{figure*}

In this section, we present the amplification technique used in our setup. We first describe its principle (Fig.~\ref{princsetup}) and discuss its expected signal-to-noise ratio. We then present the complete apparatus (Fig.~\ref{expsetup}).
\subsection{Principle of the setup}

\begin{figure}[!htph]
\centering\includegraphics[width=0.3\textwidth]{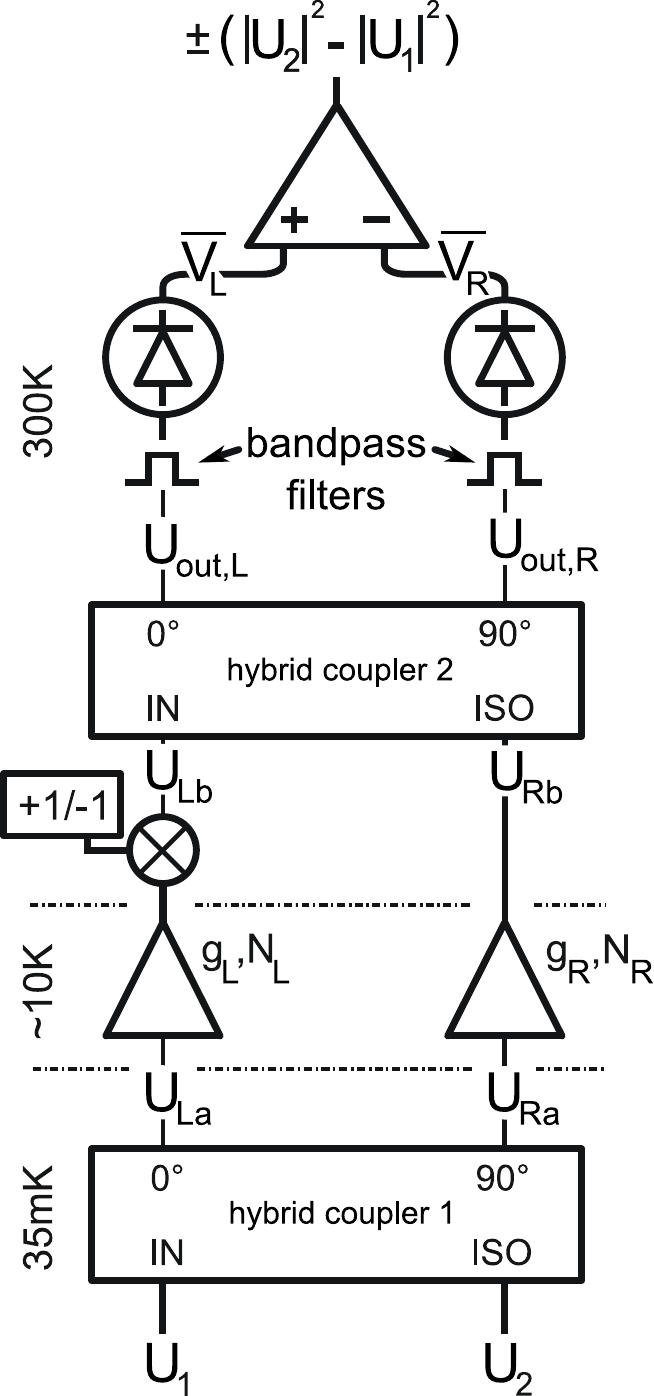}
\caption{Principle of the modulated double balanced amplifier setup. The setup measures the difference between the two input noise powers, with a $\pm1$ factor given by the modulation.
}
\label{princsetup}
\end{figure}

We use a modulated double balanced amplifier scheme (see Fig.~\ref{princsetup}) to amplify the noise of the sample. The balanced amplifier\cite{EngelbrechtIEEE} is widely used in cellular phone applications as well as in astrophysics, for downconverted millimeter radiation in recent Cosmic Microwaves Background detection\cite{Gervasi2003}, and particle physics to detect halo axions\cite{AsztalosPRD}; it can be seen as the microwave analog of a Mach-Zehnder interferometer. Its key elements are the 90-degrees hybrid couplers \cite{simons}, which act as the beam splitters in the interferometer. When the gains and phases acquired in both arms of the interferometer are equals, the signal in the first input $IN$ (resp. second input $ISO$) of the interferometer is amplified and entirely transmitted to the second output $90^{\circ}$ (resp. first output $0^{\circ}$). On the other hand, the noise of each amplifier in the inner arms is evenly distributed between the two outputs of the interferometer. As a result, when one measures the difference between the interferometer's output powers, the noise of the amplifiers vanishes and only the difference between the two input signals powers remains. In addition, when a $\pi$-phase modulator is inserted in one arm inside the interferometer, one can alternatively swap the interferometer's outputs for the signal, hence alternatively change the sign of the difference between the two input signals powers while leaving the noise of the amplifiers unchanged. This allows to completely remove the amplifiers noise in a lock-in detection.

 The $90^{\circ}$ hybrid coupler is a four ports microwave component with a S-parameters matrix $S$ between the complex amplitude of its two inputs $(IN,ISO)$ and its two outputs $(0^{\circ},90^{\circ})$ given by:

\begin{eqnarray}
S=\frac{1}{\sqrt{2}}\left(
          \begin{array}{cc}
            1 &  i\\
            i &  1\\
          \end{array}
        \right)
 \label{eq:hybridcoupler}
\end{eqnarray}

Each one of the two inner arms of the interferometer includes an amplifier with a gain $g_{i}$ and a noise $N_{i}$. The gain $g_{i}$ includes the phase difference acquired by the signal over the arm length. The left arm also includes a $\pi$-phase modulator, which multiplies the signal by a factor $\pm1$ according to the sign of the driving voltage. When the driving voltage is a low-frequency square (here, $2.7kHz$), the signal in the left arm $U_{Lb}$ periodically switches between $U_{Lb}$ and $-U_{Lb}$. The output signals of the interferometer, obtained after recombination of the left arm and right arm's signals on the second hybrid coupler, are filtered and applied to two square law detectors which measure the average power with an integration time of $0.1\mu s$. Finally, the measured difference between the two output powers is averaged over a long time $T_0$ to achieve the requested noise power resolution.
Let us first consider that the $ISO$ input signal $U_{2}$ is zero. We will show that since the setup is symmetric for the two inputs, one can easily deduce the result for two finite input signals. When the $IN$-input monochromatic signal with a complex amplitude $U_{1}$ is split by the first hybrid coupler, Eq. (\ref{eq:hybridcoupler}) gives:

\begin{eqnarray}
\left\{
          \begin{array}{c}
            U_{La}=\frac{1}{\sqrt{2}}U_{1}\\
            U_{Ra}=\frac{i}{\sqrt{2}}U_{1}\\
          \end{array}
        \right.
 \label{eq:hcoupler1}
\end{eqnarray}

Here, $U_{La}$ (resp. $U_{Ra}$) is the complex amplitude of the signal in the left (resp. right) inner arm of the interferometer, before amplification. After amplification, the signals become:

\begin{eqnarray}
\left\{
          \begin{array}{c}
            U_{Lb}=\pm(\frac{1}{\sqrt{2}}g_{L}U_{1}+N_{L})\\
            U_{Rb}=\frac{i}{\sqrt{2}}g_{R}U_{1}+N_{R}\\
          \end{array}
        \right.
 \label{eq:amplgain}
\end{eqnarray}

The $\pm 1$ factor in $U_{Lb}$ is given by the $\pi$-phase modulator. The signals are then recombined on the second hybrid coupler:

\begin{eqnarray}
\left\{
          \begin{array}{c}
            U_{out,L}=\frac{1}{\sqrt{2}}\left(\frac{1}{\sqrt{2}}(\pm g_{L}-g_{R})U_{1}\pm N_{L}+iN_{R}\right)\\
            U_{out,R}=\frac{1}{\sqrt{2}}\left(\frac{i}{\sqrt{2}}(\pm g_{L}+g_{R})U_{1}\pm iN_{L}+N_{R}\right)\\
          \end{array}
        \right.
 \label{eq:hcoupler2}
\end{eqnarray}

When the interferometer is perfectly balanced, the gains and phase differences across the inner arms are equal, giving $g_L = g_R=g$. The prefactor of $U_1$ in $U_{out,L}$ (resp. $U_{out,R}$) is then equal to $g (\pm 1-1)/2$ (resp. $g (\pm 1+1)/2$): the signal is entirely transmitted to only one output at a time, and periodically switched between the two outputs. 
The square law detectors measure the average power of the filtered signals $\overline{V_i} \propto\overline{|U_{out,i}|^2}$ over the filter bandwidth $\Delta f$:

\begin{eqnarray}
\left\{
          \begin{array}{c}
            \overline{V_L}=\frac{\alpha_1}{2}\left(\frac{|g|^2}{2}(\pm 1-1)^2 \overline{|U_{1}|^2}+ \overline{|N_{L}|^2}+\overline{|N_{R}|^2}\right)\\
            \overline{V_R}=\frac{\alpha_2}{2}\left(\frac{|g|^2}{2}(\pm 1+1)^2 \overline{|U_{1}|^2}+ \overline{|N_{L}|^2}+\overline{|N_{R}|^2}\right)\\
          \end{array}
        \right.
 \label{eq:quaddetectors}
\end{eqnarray}
$\alpha_i$ is the power to voltage conversion factor of the quadratic detectors; it includes amplification/attenuation factors in the output arms of the setup. Eq. (\ref{eq:quaddetectors}) assumes that $U_1$, $N_L$ and $N_R$ are independent, so that all correlation terms such as $\overline{U^{\ast}_{1} N_L}$, $\overline{U^{\ast}_{1} N_R}$ or $\overline{N^{\ast}_{L} N_R}$ vanish. As $N_L$ and $N_R$ have equal contributions in both outputs, they vanish in the final subtraction $\overline{V_L}-\overline{V_R}$ if $\alpha_1=\alpha_2=\alpha$. This gives:

\begin{eqnarray}
            \overline{V_{meas}}=\mp\alpha|g|^2 \overline{|U_{1}|^2}          
 \label{eq:1inpresult}
\end{eqnarray}

The measured output voltage is therefore a square signal, with a frequency $f=2.7kHz$ and an amplitude $V_{meas}=\alpha |g|^2 \overline{|U_{1}|^2}$, that can be detected with conventional lock-in measurement techniques so as to make the measurement insensitive to low-frequency variations of the amplification parameters, thus greatly enhancing the stability of the device. The contribution of a signal $U_2$ on the second input $ISO$ of the interferometer can easily be included:
\begin{eqnarray}
            \overline{V_{meas}}=\pm\alpha|g|^2 \left(\overline{|U_{2}|^2}  - \overline{|U_{1}|^2} \right)        
 \label{eq:2inpresult}
\end{eqnarray}

 One can generalize this formula to non-monochromatic input signals with current power spectral densities $S_{1,2}(f)$, meaning that the result has to be integrated over a finite bandwidth. We finally obtain:

\begin{eqnarray}
            \overline{V_{meas}}=\pm\alpha \int^{\infty}_{0}|\chi(f)g(f)|^2 (S_{2}(f)-S_{1}(f))df          
 \label{eq:2inpresultf}
\end{eqnarray}
where $\chi(f)$ is the filter function of each output arm of the device, ideally given by a square window with a bandwidth $\Delta f$ and equal for both arms. The setup therefore measures the difference of the power spectral densities of the two inputs. As described in the next section, we connect the first input to the sample output, and the second input to a load with a fixed temperature; interestingly, this differential setup can be used to measure the noise difference between two samples, or between two distinct ports of the same sample, leading to cross-spectrum measurements.

\begin{figure}[!htph]
\centering\includegraphics[width=0.4\textwidth]{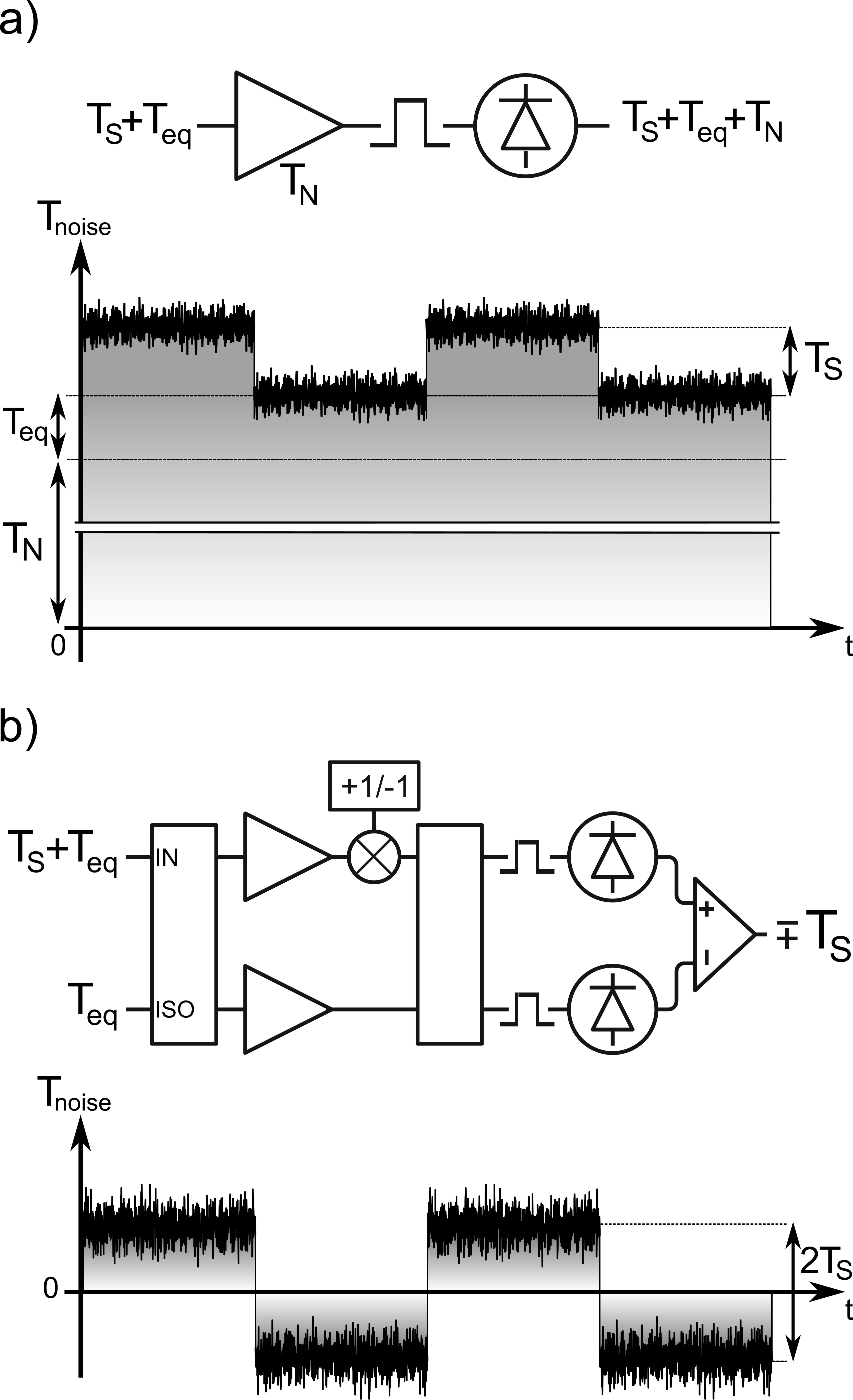}
\caption{a) Direct amplification technique: the signal is amplified, filtered and applied to the square law detector, measuring the sum of the noise temperature of the signal and the measurement load $T_S+T_{eq}$ and the noise temperature of the amplifier $T_N$. Below is a schematic representation of the measured lock-in voltage as a function of time: the value of the lock-in voltage alternatively switches between $T_{eq}+T_N$ and $T_S+T_{eq}+T_N$. The peak-peak amplitude of the detected square voltage is equal to $T_S$. b) Our setup detects and modulates the difference between the two input noises $T_S+T_{eq}$ and $T_{eq}$, that is $\mp T_{S}$. The lock-in voltage is then centered on zero while its peak-peak amplitude is equal to $2T_S$. The standard deviation is however $\sqrt{2}$ times larger in our setup.
}
\label{SNR}
\end{figure}

We shall now discuss the advantages of this setup compared to a direct amplification technique, as described in Fig.~\ref{SNR}a, where the noise of the sample is directly amplified, filtered and measured on a square law detector. For a direct comparison with the amplifiers noise temperature, we express the current fluctuations of the input signals $S_{I}$ in terms of a noise temperature $T_{S_I}$: 
\begin{eqnarray}
            Z_0 S_{I}=4 k_B T_{S_I}          
 \label{eq:noisetemp}
\end{eqnarray}
where $Z_0$ is the load impedance (generally $50\Omega$ for microwave circuits), and $k_B$ the Boltzmann constant. In each case, the sample emits a noise $T_S$, and is connected to the measurement load $Z_0$, which itself emits an equilibrium noise $T_{eq}$. In Fig.~\ref{SNR}a, the amplification adds a noise $T_N \ll T_{eq}$ (typically, $T_N \approx 7K$ and $T_{eq}\approx 30mK$) to the signal $T_S+T_{eq}$: the measured signal is then proportional to the sum $T_S+T_N+T_{eq}$. In order to extract $T_S$, one usually removes $T_N+T_{eq}$ by periodically switching on and off $T_S$ while performing a lock-in detection. In this case, the low-frequency output voltage is a square signal with an offset $T_N+T_{eq}+T_S /2$ and an amplitude $T_S /2$. If the amplifier's noise $T_N$ is Gaussian\cite{BendatPiersol}, the signal-to-noise ratio is then equal to $(S/N)_{direct}=(T_S/2T_N)\sqrt{\Delta ft_{meas}}$, where $\Delta f$ is the bandwidth of the filter, and $t_{meas}$ the measurement time. This expression can be compared to the signal-to-noise ratio calculated for our setup, see Fig~\ref{SNR}b: the sample and the measurement load are connected to the $IN$ input, and a $Z_0$ load is connected to the $ISO$ input. The noise temperature on the $IN$ input is therefore equal to $T_S+T_{eq}$, while the noise on the $ISO$ input is equal to $T_{eq}$. Our setup detects the difference between the two input noises, that is $\pm T_S$. The low-frequency output voltage is therefore a square signal with an amplitude $T_S$ and no offset. The suppression of the noise offset due to the amplifiers greatly enhances the stability of the setup, since one is no more sensitive to variations of the amplifiers noise, which are usually much larger than the signal $T_S$. This result is illustrated by the graphs in Fig.~\ref{SNR}, which represent the measured lock-in voltage as a function of time for the direct amplification scheme (a) and our setup (b).

In our setup, the standard deviations of the two amplifier's noises add, while the noise offset due to the amplifiers is zero after the final subtraction. The standard deviation of the amplification noise in our setup is then $\sqrt{2}$ times the fluctuations of a single amplifier. 
However, since the amplitude of the measured noise in lock-in detection is double in our setup, the signal-to-noise ratio is still larger than in the direct amplification scheme, and given by:

\begin{eqnarray}
            \left(\frac{S}{N}\right)_{setup}=\frac{T_S}{\sqrt{2}T_N}\sqrt{\Delta ft_{meas}}          
 \label{eq:signaltonoise}
\end{eqnarray}

\begin{figure}[!htph]
\centering\includegraphics[width=0.4\textwidth]{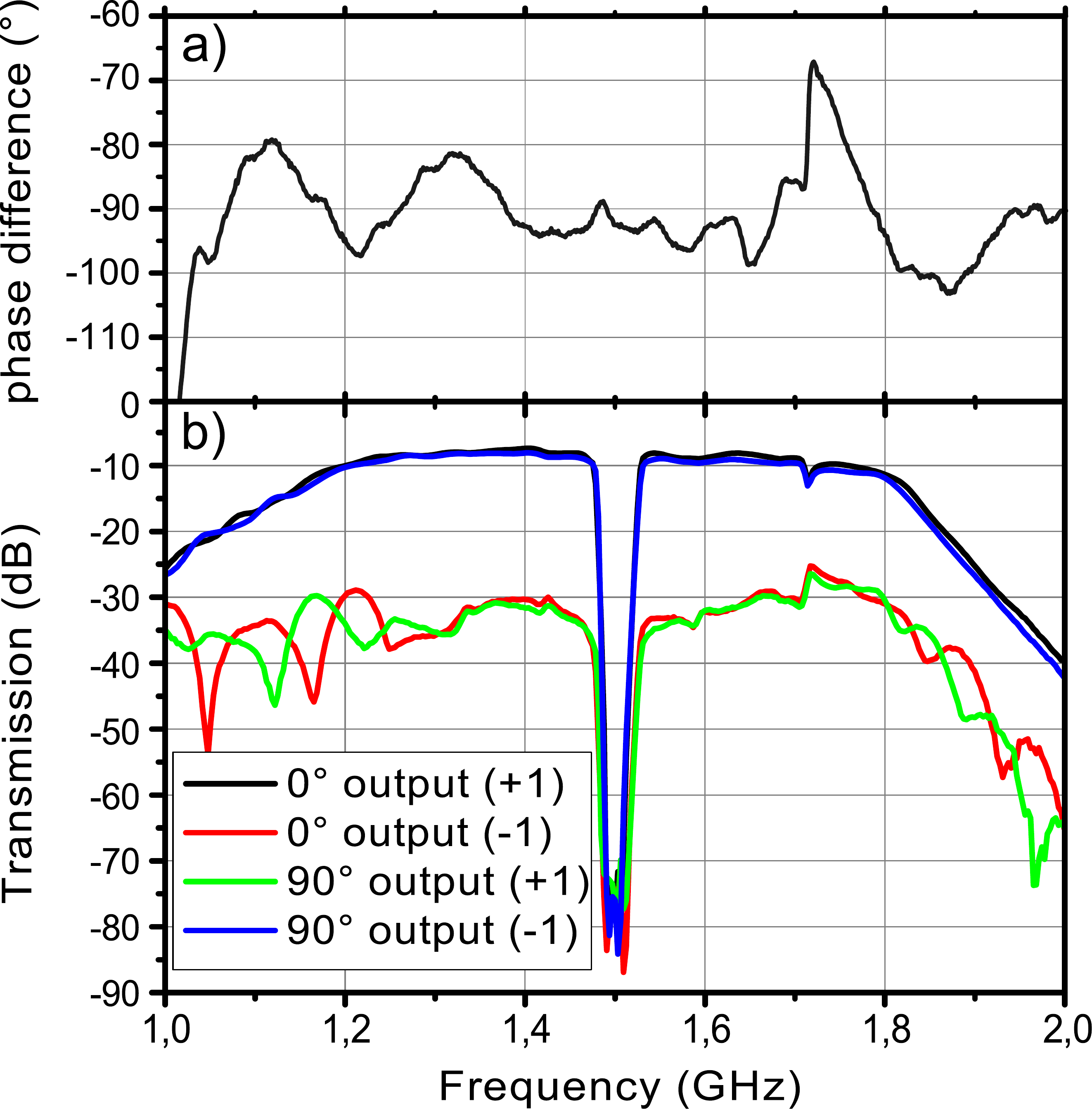}
\caption{a) Phase difference between the two inner arms of the device as a function of the frequency. The signals of the two arms are measured just before the second hybrid coupler. b) Transmission between the input of the refrigerator and the two output arms of the setup (just before the square law detectors) for a positive $(+1)$ and negative $(-1)$ DC voltage on the modulator. The $1.5GHz$ carrier is suppressed by more than $60dB$.
}
\label{exptuning}
\end{figure}
For a given signal-to-noise ratio, our setup therefore allows measurements twice as fast as a direct amplification technique. However, Eq. (\ref{eq:signaltonoise}) stands for a perfectly balanced setup. Using the same calculations for a non-balanced setup, we expect the signal-to-noise ratio to be diminished by $5\%$ for a 3dB gain difference between the output arms, and the measurement time to be increased by $3\%$ for a $10^{\circ}$ phase difference between the two inner arms.
Furthermore, the suppression of the noise offset due to the amplifiers greatly enhances stability, since the slow variations of $T_N$ are automatically compensated.

\begin{figure*}[!htph]
\centering\includegraphics[width=0.80\textwidth]{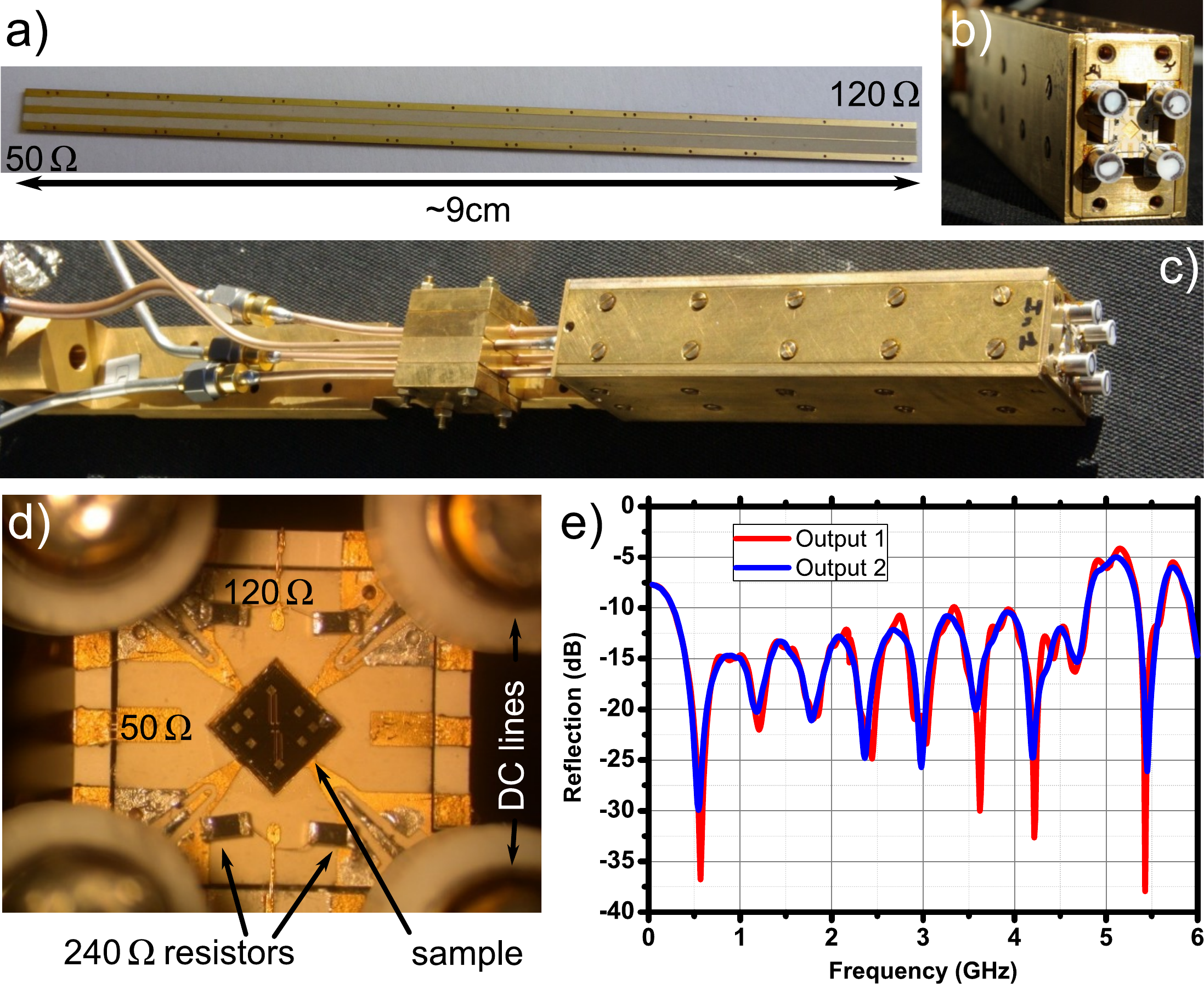}
\caption{a) $120\Omega-50\Omega$ transformer line: the coplanar waveguide is built on a $TMM10$ substrate for low-temperature performance. The width of the center conductor is $0.66mm$ for the $50\Omega$ port, and $0.075mm$ for the $120\Omega$ port. b) and c) Pictures of the 4-microwave ports sample holder. The $50\Omega$ lines and the transformer lines are encased in the four sides of the sample holder. d) Zoom on the center part of the sample holder; the size of the sample is $2mm\times 2mm$. e) Reflection on the $50\Omega$ port of the two transformer lines as a function of the frequency, measured in liquid nitrogen.
}
\label{transfrm}
\end{figure*}
The modulated double balanced amplifier technique is thus expected to increase the stability and sensitivity for high frequency noise measurements, while being relatively robust to imperfections in the setup.

\subsection{Implementation and calibration}

The implemented setup is shown in Fig.~\ref{expsetup}. Two cryogenic amplifiers (\textsl{MiteQ AFS3-02000400-08-CR-4}) are used with a noise temperature of about $7K$ when thermalized at $10K$ in Helium vapor, and an extended bandwidth of $1-4GHz$; these amplifiers can present a noise temperature as low as $3.5K$ when thermalized in a pumped bath at $1.8K$ \cite{Chincarini}.  Up to the dilution refrigerator's outputs, the setup is wired with \textsl{UT-85 SS} semirigid cryogenic microwave cables for an optimized thermalization. We also protect the sample from the back-action noise of the amplifiers using \textsl{Pamtech LTC 1384K4} cryogenic circulators whose $50\Omega$ loads are thermalized to the mixing chamber of the dilution fridge to reduce the background thermal noise. These optional circulators restrict the bandwidth of the whole setup to $1.2-1.8GHz$. The lengths of the inner arms are matched using a phase shifter to tune the length of the second arm. $3dB$ attenuators are regularly placed in between room temperature parts of the setup to suppress multiple reflections between the components; the $6dB$ attenuator in the first inner arm is used to balance the gain difference between amplifiers $A1$ and $A2$. We insert a $\pi$-phase modulator (\textit{Miteq BMA0104LA1MD}) in each inner arm to symmetrize the insertion losses and phase shifts ($~90^\circ$); however, we modulate only the signal in the first arm, feeding the first modulator with a $2.7kHz$ square voltage through a $600\Omega$ load while the second modulator is fed with a constant current. After recombination on the second hybrid coupler, the signals are filtered in the $1.2-1.8GHz$ band. We use a $1.5GHz$ excitation voltage to drive the sample out of equilibrium. This adds a $1.5GHz$ carrier frequency to the signal, containing informations on the conductance of the sample, as well as a parasitic signal. We derive a portion of the signal in the second output arm using a $6dB$ splitter (compensated by a $6dB$ attenuator in the first output arm) and detect the in-phase and out-of-phase parts of the carrier frequency with a homodyne detection. In analogy with optics, we use a $90^{\circ}$ hybrid coupler as a beam splitter and multiply the $0^{\circ}$ and $90^{\circ}$ outputs by a $1.5GHz$ local oscillator. The result of the multiplication of the $0^{\circ}$ (resp. $90^{\circ}$) output yields a zero-frequency part proportional to the in-phase (resp. out-of-phase) part of the carrier frequency. When the modulation is turned on, the carrier frequency is switched between the two output arms; therefore, the homodyne signals are $2.7kHz$ square voltages switching between zero and a value proportional to the quadrature components of the carrier frequency, and are detected with lock-in techniques. 
In the noise measurement part of the setup, the $1.5GHz$ carrier frequency is removed ($-70dB$) with $1.5GHz$ notch filters. The noises in the two output arms are subtracted with a \textit{NF LI75-A} low frequency differential amplifier.

We have tuned the setup to optimize the phase and gain balance in the inner arms, as well as the gain balance in the output arms. The latter is done by inserting a variable attenuator, set to $0dB$, in the second output arm (the insertion loss of the attenuator compensates the gain difference in the arms). In order to characterize the gain and the phase balance, we use a vector network analyzer to measure the transmission between the first input of the setup with a $90dB$ attenuation, and each one of the two inner arms just before the second hybrid coupler (Fig.~\ref{exptuning}a), or each one of the two output arms just before the square law detectors (Fig.~\ref{exptuning}b). The second input of the setup is connected to a $50\Omega$ load thermalized to the mixing chamber, and the first modulator (\textit{mod1}) is fed with a constant (positive or negative) voltage to study both situations. 

The results of the tuning are shown in Fig.~\ref{exptuning}. The phase balance is achieved within $\pm5^{\circ}$ in the $1.2-1.8GHz$ bandwidth, which only degrades the signal-to-noise ratio by a few percents. As a result, the test signal is transmitted to only one output, with less than $1\%$ of the power transmitted to the other output. This $20dB$ difference between the two transmissions compares favorably with standard isolation values in commercial-grade microwave components. The amplification and filtering are identical (to less than $1dB$) for both outputs. The $1.7GHz$ peak in the phase balance, due to the cryogenic circulators, causes a decrease of the transmission difference to $~15dB$, which is still within acceptable bounds. 

We have calibrated the setup by replacing the thermalized $50\Omega$ load connected to the second input with a variable temperature $50\Omega$ load, which acts as a tunable thermal noise source. The temperature of the load is measured with a calibrated $RuO_2$ resistance. We use a series of SMA connectors to thermically decouple the load from the mixing chamber. We obtain a calibration between input temperature difference $\Delta T$ and the amplitude of the measured $2.7kHz$ voltage: $V_{lock-in}=1.37\times 10^{-5}(\pm 5\%) \Delta T$.

\section{Quarter-wave impedance transformer}

For given current fluctuations, one can increase the equivalent noise temperature by increasing the load impedance $Z_0$ in Eq. (\ref{eq:noisetemp}). However, since a vast majority of commercial microwave components are $50\Omega$-adapted, one needs to transform the impedance seen by the sample from the increased $Z_0$ (in our case, $Z_0=120\Omega$) to $50\Omega$ while keeping a large bandwidth. This can be achieved by using a quarter wave impedance transformer\cite{pozar,EZB}, which consists of a series of coplanar waveguides with gradually changing impedances. Every coplanar section has the same length, given by the quarter of the wavelength at center frequency. Depending on the series of impedances, one can either optimize the gain flatness or the total bandwidth. 

\begin{figure}[!htph]
\centering\includegraphics[width=0.45\textwidth]{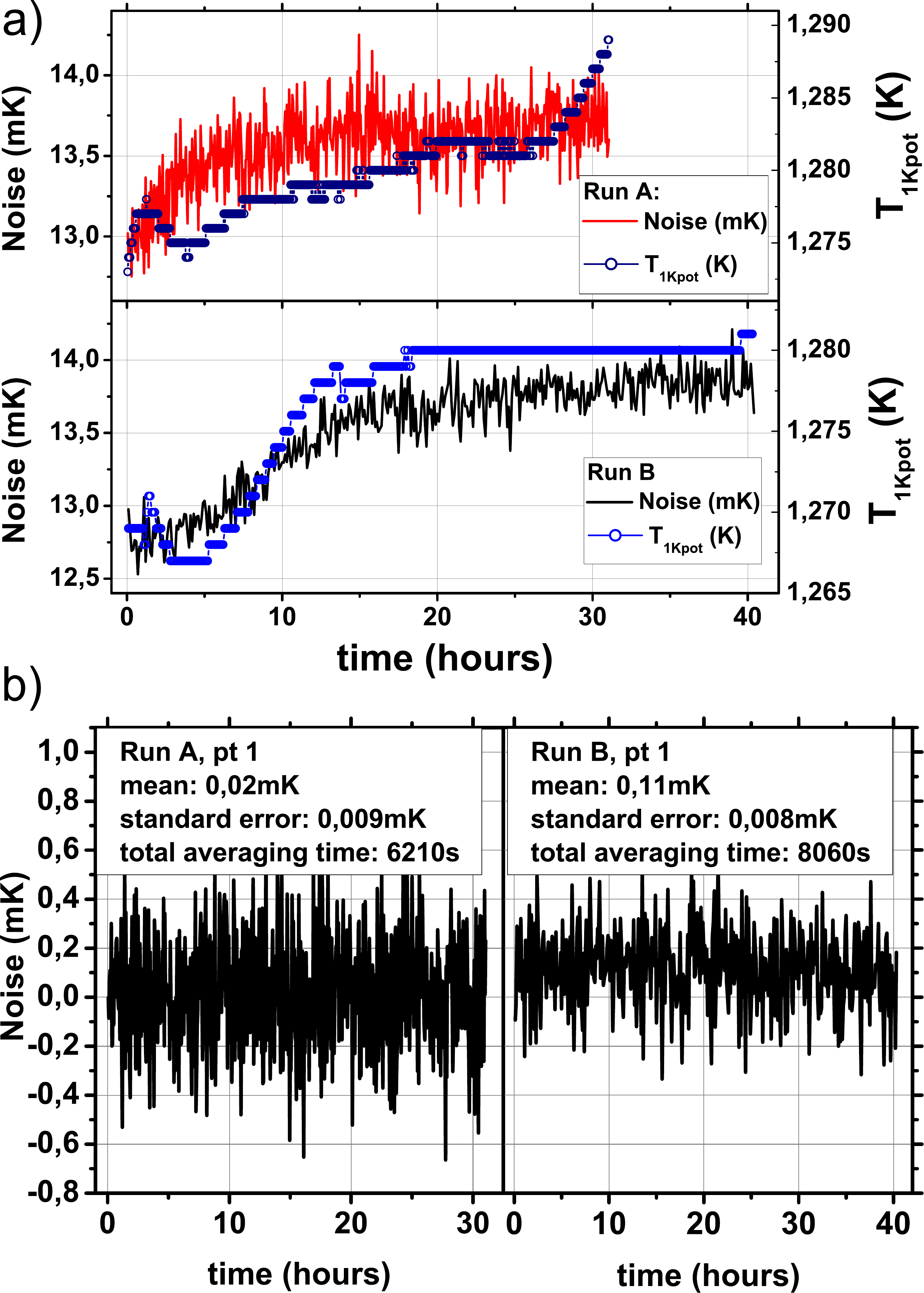}
\caption{Operation of the device: a) stability of the device for two non-consecutive runs: measured noise (line) and temperature of the 1K pot (circles) as a function of time. Noise data  for both graphs are measured for the same reference gate voltage of the sample $Vg_{ref}$. The averaging time per point is $10\;s$ for run A, and $20\;s$ for run B. b) datasets obtained after subtraction of noise at the reference gate voltage ($T_S(Vg_{1})-T_S(Vg_{ref(1)})$) for the first points of both runs (measured at different gate voltages). The dataset for run A presents a significantly larger standard deviation due to the shorter averaging time per point.
}
\label{expoperation}
\end{figure}

We designed an 8-sections Chebychev (equal ripple) $120\Omega-50\Omega$ transformer\cite{MWdesign} (see Fig.~\ref{transfrm}a), allowing a large bandwidth ($0.5-4.5GHz$). The $120\Omega$ port is shunted by two $240\Omega$ $NiCr$ resistors in parallel (see Fig.~\ref{transfrm}d) to avoid back-reflection of the noise of the measurement setup on the sample connected in parallel to the resistors, thus acting as a $120\Omega$-adapted current source (we neglect the influence of the sample's impedance, of a few $K\Omega$). We have taken into account the parasitic capacitances of the resistors (typically $0.03pF$) and the sample ($~0.06pF$) by changing the length of each section to optimize the transmission of the device. We use a 4 microwave ports geometry for the sample holder; the two input ports are $50\Omega$-adapted while each output port includes an impedance transformer. Both input and output lines are coplanar waveguides built on a TMM10 substrate, and encased in a copper sample holder (Fig.~\ref{transfrm}b and c) thermalized to the mixing chamber of the dilution refrigerator. We have characterized the frequency response of the transformer by measuring the reflection of the $120\Omega$ port as a function of the frequency (see Fig.~\ref{transfrm}e). We find a reflection of $~15dB$ at $77K$, which is comparable to the reflection factors in commercial microwave components. This corresponds to a power transmission through the transformer of $97\%$. The use of the transformer allows to increase the power spectral density of the measured signals by a factor $2.4$. For a noise temperature of the amplifiers of about $7K$ and a $120\Omega$ measurement load, Eq. (\ref{eq:signaltonoise}) gives an expected sensitivity of $2\times10^{-28}A^{2}/Hz/\sqrt{Hz}$.

\section{Operation of the setup}

We combine the effects of the quarter-wave impedance transformer and the modulated double-balanced amplifier to increase the signal of our sample, and measure it over extended periods of time with a large stability. In a standard noise measurement of a mesoscopic sample, one usually measures the noise $T_S(Vg)$ as a function of the device parameters, which can be tuned using one (or more) gate voltage $Vg$. Since the amplification parameters as well as the temperatures of the different stages of the dilution refrigerator can vary over the usual averaging times (about 1 hour per point), we perform repeated short measurements of the noise for a few (typically 5) gate voltages $Vg_{1,..,5}$ and a reference gate voltage $Vg_{ref}$ which defines the zero of the measured noise. We thus measure the excess noise compared to a reference operating point of the sample. Since the measurement device is highly sensitive, one has to make sure that the temperature difference between the $120\Omega$ load connected to the sample and the load connected to the second input of the interferometer varies as slowly as possible. We connect the $120\Omega$ load of the second impedance transformer built on the sample holder (see Fig.~\ref{transfrm}d) to the second input of the interferometer to keep the same thermal environment for the two loads, as well as reduce the offset due to the noise temperature difference between a $120\Omega$ and a $50\Omega$ load. We also stabilize the temperature of the mixing chamber within less than a milliKelvin using the \textit{femtopower} temperature regulation provided with Oxford Kelvinox refrigerators.

\begin{figure}[!htph]
\centering\includegraphics[width=0.5\textwidth]{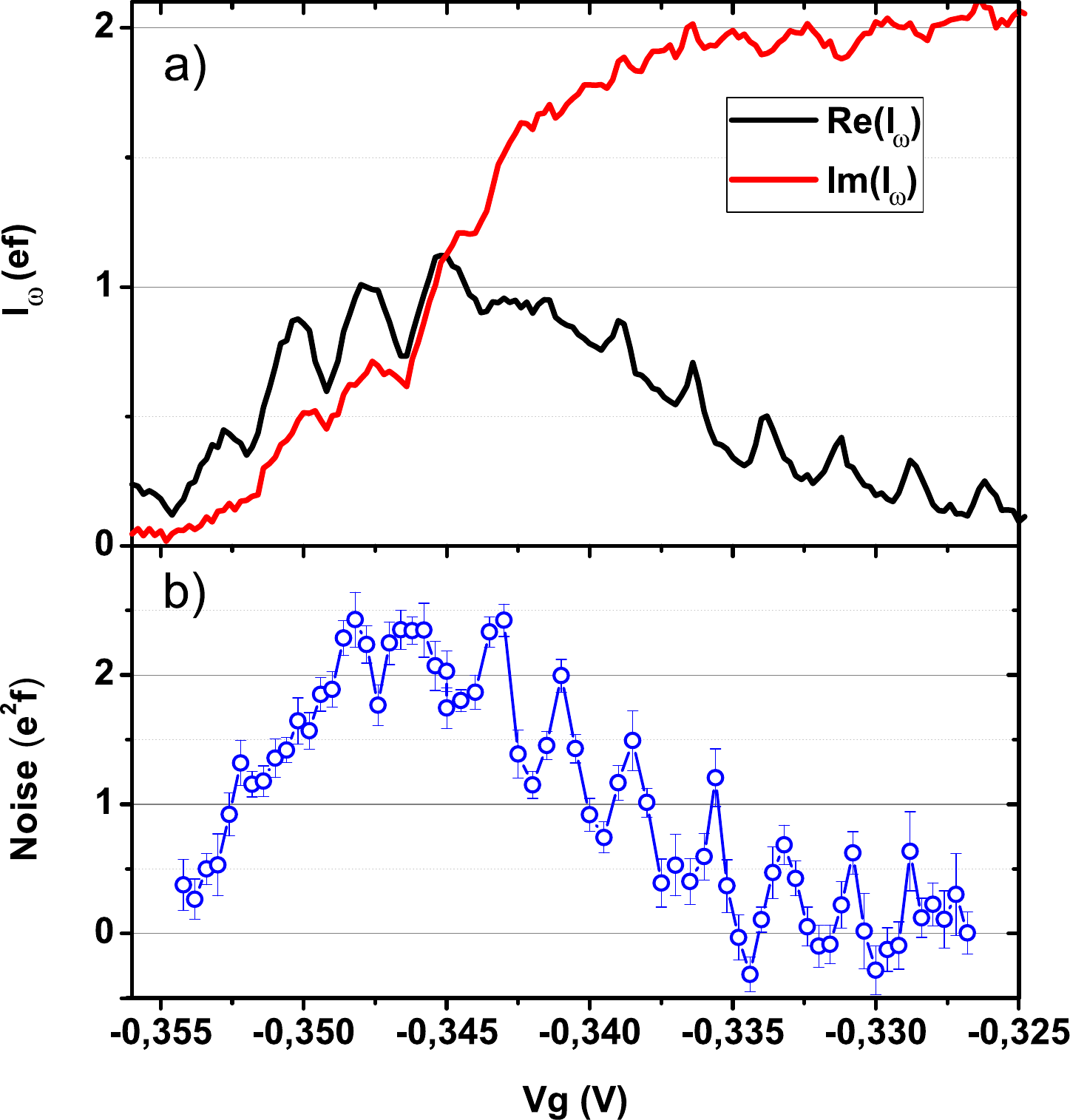}
\caption{Average ac current and noise of a single electron source measured by our setup: a) in-phase and out-of-phase parts of the average ac current emitted by the source at $f=1.5GHz$, as a function of the gate voltage $Vg$ controlling the coupling between the dot and the electron gas. In this case, the signal ($I_\omega\propto ef\approx 0.25nA$) can easily be measured in less than a second. b) Current autocorrelations measured with our setup as a function of $Vg$. The error bars demonstrate that our setup is well-suited for precise measurements of noise spectral densities lower than $4\times 10^{-29} A^2/Hz$.
}
\label{expnoise}
\end{figure}

A typical operation of the device is presented in Fig.~\ref{expoperation}a: we measure the noise for each of the 5 gate voltages $T_S(Vg_{1,..,5})$ during a short time ($10\;s$ for run A, $20\;s$ for run B). We systematically measure the noise for the reference gate voltage $T_S(Vg_{ref})$ after each gate voltage, thus creating a sequence composed of 10 short measurements ($T_S(Vg_{1})$, $T_S(Vg_{ref(1)})$, $T_S(Vg_{2})$, $T_S(Vg_{ref(2)})$, and so on), which we repeat a large number of times (621 for run A, 403 for run B). The total averaging time for each point is therefore at least ten times shorter than the total measurement time; a significant portion (one third for run A) of the total measurement time is spent in setting the gate voltage to its different values. We then remove the long-time variations of the signal due to slow temperature changes in the dilution refrigerator (such as the 1K pot temperature plotted in Fig.~\ref{expoperation}a) by calculating the difference between the traces obtained for each gate voltage and their respective reference: $T_S(Vg_{i})-T_S(Vg_{ref(i)})$. We finally calculate the mean value of each set of data such as the two presented in Fig.~\ref{expoperation}b to obtain the noise, while the sensitivity of the measurement is given by the standard error.  Fig.~\ref{expoperation}b demonstrates a noise temperature resolution $T_{res}$ of less than $10\mu K$ (\textit{i.e.} $4.6\times 10^{-30} A^2/Hz$) in about 2 hours; this gives a sensitivity $s=T_{res}\sqrt{t_{mes}}$ equal to  $0.71mK/\sqrt{Hz}$, \textit{i.e.} $3.3\times 10^{-28} A^2/Hz/\sqrt{Hz}$ on a $120\Omega$ load. This value of the sensitivity is larger than the theoretical value; however, one has to consider the fact that the noise values are obtained after subtraction of a reference noise, hence multiplying the standard error by a factor $\sqrt{2}$. The calculated effective sensitivity of the measurement is thus $\sqrt{2}$ times larger than the sensitivity of the setup, which is then equal to $2.3\times 10^{-28} A^2/Hz/\sqrt{Hz}$. This value is close to the theoretical sensitivity, demonstrating the good implementation of the device, and a large enough stability to perform measurements averaged over several hours. 

We finally present an example of the measurements made possible by our setup. We have used our setup to fully characterize a single electron source described in Ref.\cite{Fevescience} consisting of a submicron dot coupled to a 2-dimension electron gas through a tunable barrier. By measuring the in-phase and out-of-phase parts of the average ac current ($f=1.5GHz$) emitted by a single electron source (Fig.~\ref{expnoise}a), as well as the current fluctuations in the $1.2-1.8GHz$ bandwidth (Fig.~\ref{expnoise}b), we have demonstrated the triggered single charge emission by the source\cite{MaheNoise}. The noise data have been obtained in about five days, each data point being measured in a total of 40 minutes using the measurement procedure described in the beginning of this section.

\section{Conclusion}

We have developed a highly sensitive microwave noise measurement setup able to detect fluctuations generated at milliKelvin temperatures over a large bandwidth. The sensitivity of the implemented setup is close to the theoretical sensitivity, demonstrating its robustness  to imperfections. The dual-inputs geometry allows comparative noise measurements such as schemes involving two different samples, or a multiterminal sample. The setup can also be modified to measure high frequency cross-correlations between the two inputs by inserting a $180^{\circ}$ hybrid coupler before the square law detectors, thus measuring $|(U_1-U_2)^2-(U_1+U_2)^2|$. We have recently used the device to measure the autocorrelations of the current generated by a single electron source, thus demonstrating that it is the proper tool for probing the outcomes of gigahertz single charge electronics and electron quantum optics experiments.

\section{Acknowledgments}
D. C. G. thanks Renzo Vaccarone for providing data on cryogenic \textit{MiteQ} amplifiers.\\

\noindent $^{\dag}$also at Service de Physique de l'Etat Condens\'e, CEA Saclay, 91191 Gif-sur-Yvette, France.

\noindent $^{\ddag}$Electronic address :
francois.parmentier@lpa.ens.fr

\end{document}